\documentclass[epj]{svjour}
\usepackage{latexsym}
\usepackage{graphics}
\usepackage{amsmath}
\usepackage{graphicx}
\usepackage{amssymb} 
\usepackage{graphics}
\usepackage{psfrag}
\usepackage{float}
\usepackage{exscale}
\usepackage{array}

\begin{document}
\hyphenation{nanomechanical nano-mech-an-i-cal}
\title{Noise color and asymmetry in stochastic resonance with silicon nanomechanical resonators}
\author{Tyler Dunn \and Diego N. Guerra \and Pritiraj Mohanty\thanks{\email{mohanty@physics.bu.edu}}}
\institute{Department of Physics, Boston University, 590 Commonwealth Avenue, Boston, Massachusetts 02215, USA}
\date{\today}

\abstract{
Stochastic resonance with white noise has been well established as a potential signal amplification mechanism in nanomechanical two-state systems.  While white noise represents the archetypal stimulus for stochastic resonance, typical operating environments for nanomechanical devices often contain different classes of noise, particularly colored noise with a $1/f$ spectrum.  As a result, improved understanding of the effects of noise color will be helpful in maximizing device performance.  Here we report measurements of stochastic resonance in a silicon nanomechanical resonator using $1/f$ noise and Ornstein-Uhlenbeck noise types. Power spectral densities and residence time distributions provide insight into asymmetry of the bistable amplitude states, and the data sets suggest that $1/f^{\alpha}$ noise spectra with increasing noise color (i.e. $\alpha$) may lead to increasing asymmetry in the system, reducing the achievable amplification. Furthermore, we explore the effects of correlation time $\tau$ on stochastic resonance with the use of exponentially correlated noise. We find monotonic suppression of the spectral amplification as the correlation time increases.
 % end of PACS codes
} %end of abstract

\maketitle
\section{Introduction}\label{intro}

Although it is not surprising given the wealth of mechanical, electronic, biological and other systems which display stochastic resonance (SR) \cite{Gammaitoni-review}, the observation of noise-enhanced switching in nanoelectromechanical systems \\ (NEMS) holds promise for the possible exploitation of noise to improve device performance \cite{Badzey-SR}. In addition, the discovery of SR in NEMS has made available yet another experimental platform for the study of basic questions in the field, such as the effect of noise color. Combining these motivations, we discuss measurements of SR in a silicon nanomechanical resonator using various colored noise spectra.

Improving the model for SR in this complex application warrants additional attention as NEMS continue to emerge as a viable technology. In particular, the bistable amplitude states of a nonlinear nanomechanical resonator may see use as two-state devices such as memory elements and switches \cite{Badzey-memory,Guerra-switch}. A particle subject to white noise in a periodically-modulated double-well potential remains a powerfully intuitive model for stochastic resonance, capturing the basic dynamics in a diverse array of systems. Still, much work has been devoted to generalizing this picture. Asymmetric potentials \cite{Bartussek,Bulsara,Nikitin,Wio}, spatially extended systems \cite{Benzi}, and $``$non-conventional$"$ stochastic resonance between coexisting periodic attractors \cite{Dykman} have been among the many extensions to the original idea, with the latter two offshoots both applying to nanomechanical SR. Such derivations continue to be important in understanding stochastic resonance in actual practice, where the experimental conditions are rarely straightforward.

Yet another important and ongoing consideration is in understanding the effect of noise color on stochastic resonance \cite{Gaimmatoni-PRL/PRA,Hanggi,Kiss,Mantega,Nozaki,Fuentes,Guerra-SR}. In particular, the ubiquity of $1/f$ noise makes it a prominent factor in a wide range of settings \cite{Press}, including in the active electronic elements used to drive nanomechanical resonators \cite{Kogan}. This noise class has been used to induce SR in electronic \cite{Kiss}, neurological \cite{Nozaki}, and most recently, nanoelectromechanical systems \cite{Guerra-SR}. Although the suppression of SR persists as a common theme in these studies, to the authors knowledge, an accepted physical picture for the effect of $1/f^{\alpha}$ noise exponent is lacking. Asymmetry has been postulated as a consequence of $1/f$ noise, and evidence in this study further reinforces that idea.

Extending the study to another class of colored noise, we also observe SR with the use of exponentially correlated (Ornstein-Uhlenbeck) noise. We find that the amplification of the signal declines with increasing correlation strength of the noise, in agreement with theoretical predictions and previous experimental results \cite{Gaimmatoni-PRL/PRA,Hanggi,Mantega}.

Previous studies of SR in nanomechanical resonators have focused mainly on traditional measures, such as signal-to-noise ratio (SNR) and spectral amplification \cite{Badzey-SR,Guerra-SR,Almog}. Given the technological potential of these systems, enhancements seen in SNR and signal amplification provide obvious implications for the utility of SR in an applied setting. However, the wealth of analyses that the field has grown to embrace, including of power spectral densities and residence time distributions, afford further insights. We use these approaches here to discern the role of asymmetry on system dynamics.

\section{Experimental Methods}\label{exp}

\subsection{Device Fabrication and Characterization}\label{device}

Using e-beam lithography, resonators are patterned on the single crystal silicon layer of a silicon-on-insulator (SOI) wafer. The length $L$ and width $w$, 20 $\mu$m and 300 nm, respectively, place the in-plane resonance frequency of the doubly-clamped beam in the MHz range. A gap $d$ of 250 nm at equilibrium separates the resonator from adjacent parallel electrodes used for actuation and detection, and the device layer thickness is 500 nm. Metal contacts are deposited via thermal evaporation, giving the resonator a thickness $t$ of 550 nm. Finally, reactive ion etching (RIE) and hydrofluoric acid etching are used to define and suspend the silicon structure. 

\begin{figure}
% Use the relevant command for your figure-insertion program
% to insert the figure file.
% For example, with the option graphics use
\resizebox{\columnwidth}{!}{%
  \includegraphics{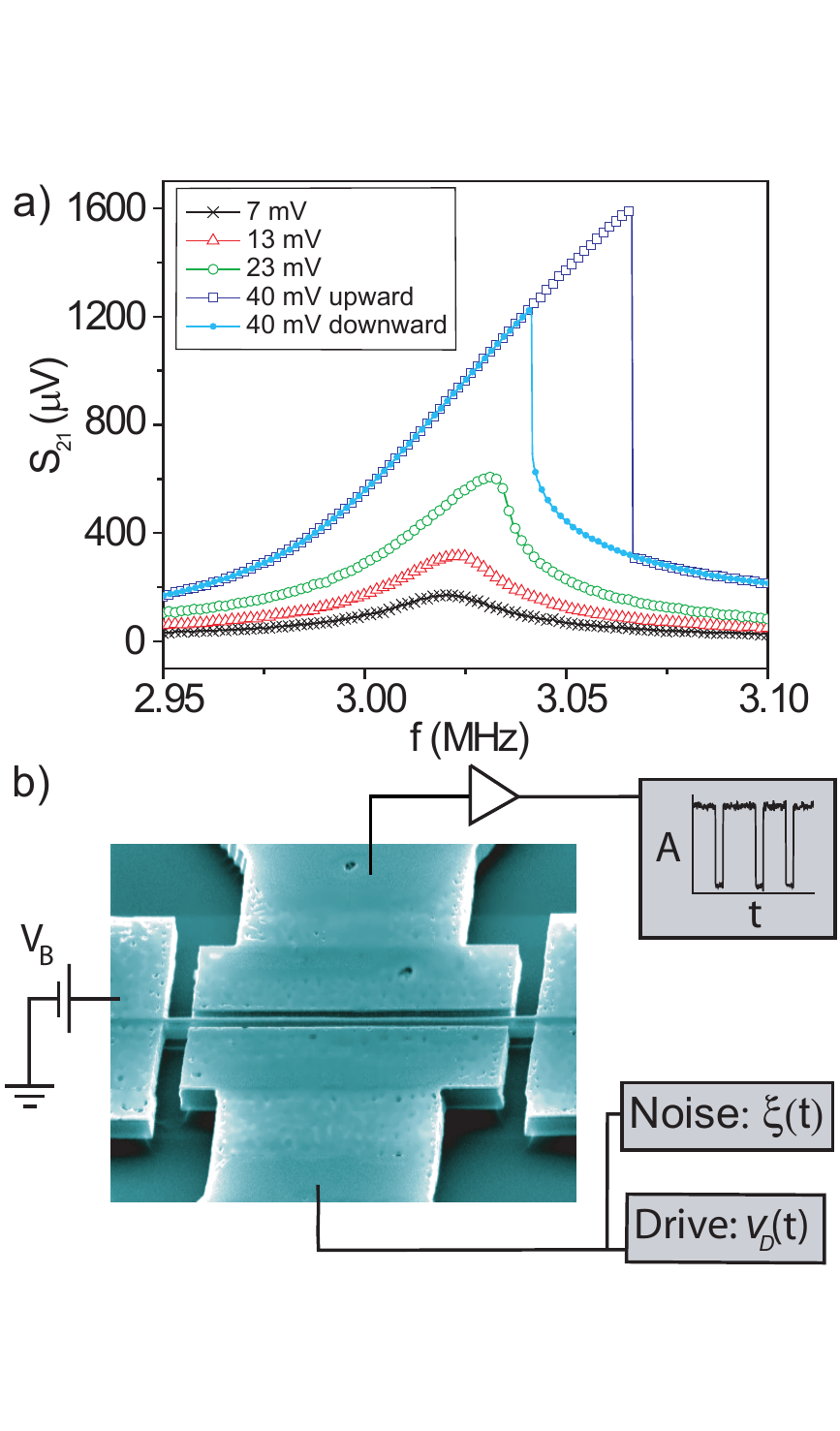} }
\caption{a) Resonator response vs. frequency at various drive levels. Beginning as a Lorentzian with resonance frequency $f_{0}$ $\approx$ 3 MHz and quality factor Q $\approx$ 50, the resonance bends to higher frequency with increasing drive voltage. Eventually, upward (empty squares) and downward (filled circles) frequency sweeps at a single drive voltage exhibit hysteresis, revealing two stable amplitude states.  b) Micrograph of the resonator with SR measurement setup. Separate sources provide the modulated drive signal $v_{D}(t)$ and the noise spectrum $\xi(t)$, while a network analyzer operating in continuous wave mode monitors the response amplitude at the drive frequency $\omega$.}
\label{setup}
\end{figure}

To actuate the beam, we utilize the standard electrostatic technique at room temperature (\cite{Guerra-switch} and references therein). With a DC bias $V_B$ charging the resonator, a high frequency excitation signal $v_{D}(t)$ applied to the drive electrode creates a capacitive force. The system can be modeled as a Duffing oscillator \cite{Nayfeh}:

\begin{equation}
\ddot{x} + \gamma\dot{x} + \omega^{2}_{0}x + k_{3}x^3 = \frac{1}{2} \frac{dC}{dx} V_B v_{D}(t)
\label{duffing}
\end{equation}

\noindent where $x$ is the effective resonator displacement, $\gamma$ is a dissipation coefficient, $\omega_{0}$ is the resonant angular frequency of oscillation, and $k_{3}$ is the non-linear coefficient. Here, the resonator and drive electrode may be accurately approximated as parallel plates, making the capacitance between them $C$ = $\epsilon_{0}$$L$$t$/($d$-$x$). In-plane oscillation of the biased resonator induces a current at the opposite electrode, which we amplify and measure with a network analyzer. The frequency response exhibits the expected Lorentzian line shape for small displacements (Fig.~\ref{setup}(a)). As drive power is increased, nonlinearity leads to spring hardening, and two stable amplitude states emerge with the onset of hysteresis. These bistable amplitude states provide a two-state system for the study of SR.

%------------------------------------------------------
%--------------Figure 1--------------------------------
%------------------------------------------------------
%------------------------------------------------------
%------------------------------------------------------
%------------------------------------------------------

\subsection{Modulation and Noise}\label{mod}

Whereas previous studies of SR in nanomechanical systems have utilized an additional periodic signal for switching \cite{Badzey-SR,Almog}, we modulate the potential by means of a phase deviation $\varphi_{0}$ in the drive signal itself. Recent experiments have shown that an abrupt phase shift can be used to induce a switch between the bistable amplitude states of a nanomechanical resonator \cite{Guerra-switch}. Here, we drive the system in the bistable region and modulate the phase of the drive with a square wave:

\begin{equation}
v_{D}(t) = v_Dcos(\omega t+\frac{\varphi _0}{2}\Theta(\Omega t))
\label{modulation}
\end{equation}

\noindent where $\Theta(\Omega t)$ represents a square wave of period $\frac{1}{\Omega}$, alternating between +1 and -1 each successive half period. The drive frequency $\omega$/$2\pi$ is a frequency in the bistable region, and $\Omega$ $<<$ $\omega$/$2\pi$ ($\Omega$ = 50 Hz for all results reported here). Expanding this expression, we see that the drive consists of two parts:

\begin{equation}
v_{D}(t) = v_D[cos(\frac{\varphi _0}{2})cos(\omega t) + \Theta(\Omega t)sin(\frac{\varphi _0}{2})sin(\omega t)].
\label{modulation2}
\end{equation}

\noindent The first term in the brackets provides simple periodic forcing. The second term, which changes phase by 180 degrees each half period of the square wave, induces switching of the resonator amplitude in synchronization with $\Theta(\Omega t)$. The magnitude of $\varphi _0$ dictates the amplitudes of these two terms, and for small angles, the amplitude of the modulation signal is proportional to $\varphi _0$. The exact effect of this modulation depends upon the specific drive conditions of the resonator, but in general, phase deviations close to $\pi$ radians induce a switch. As the magnitude of the phase deviation is dropped, the switching fidelity drops as well, and below a certain value of $\varphi_{0}$, the modulation becomes sub-threshold and the resonator remains in the given amplitude state (see, for example, Figure 3 in \cite{Guerra-switch}). In these experiments, we set the periodic modulation just below this threshold and recover coherent switching via the addition of relatively small noise intensities to the system. This places the measurements in the weak noise limit, outside the applicability of linear response approximations \cite{Shneidman}.

Colored noise spectra are digitally generated in MatLab and reproduced by a second signal generator. The signal generator implements I/Q modulation of a carrier signal, producing a voltage $\xi(t)=I(t)cos(\omega_ct)+Q(t)sin(\omega_ct)$, where $I(t)$ and $Q(t)$ are the digitally generated noise sequences, and $\omega_c$ is the angular frequency of the carrier signal. We generate
$1/f^{\alpha}$ noise by creating a sequence of 63,000 points in frequency space with an amplitude of the form $1/f^{\frac{\alpha}{2}}$ and random phase (taken from a uniform distribution $[-\pi,\pi]$). Taking the inverse Fourier transform of this set, we obtain the desired noise sequence in the time domain. In the case of Ornstein-Uhlenbeck noise, the noise sequences are generated by solving the equation:

\begin{equation}
\dot{\xi}_{OU}(t) = \frac{-1}{\tau}\xi_{OU}(t) + \frac{\sqrt{D}}{\tau}\xi_{wh}(t)
\label{OrnUhl}
\end{equation}

\noindent where $\tau$ is the correlation time, $D$ is the noise intensity and $\xi_{wh}(t)$ is a sequence of 63,000 points of Gaussian white noise generated in MatLab using the $randn()$ function.

\subsection{Measurements}\label{meas}

Bistability of the resonator extends over a relatively wide range of operable frequencies and drive voltages.  Furthermore, the drive parameters influence the relative stability of the two amplitude states \cite{Nayfeh}, affecting the symmetry of the potential describing the system.  As such, care is taken prior to measurement to symmetrize the states as closely as possible.  In the absence of noise and with the phase modulation inducing infrequent switching, we adjust the high frequency drive voltage until neither amplitude state is strongly preferred. The phase modulation is then dropped to a sub-threshold level, and noise is added within a 50 kHz band encompassing the full width at half maximum (FWHM) of the resonance. In the case of $1/f$ noise, we choose an I/Q carrier frequency at the low frequency side of the beam resonance, so that noise power spectral density (PSD) decays across the bistability region in frequency space. By contrast, we center the Lorentzian line shapes of exponentially correlated noise within the bistability region itself. 

Figure~\ref{setup}(b) displays the measurement setup. For each value of noise intensity, a network analyzer monitors the response at the drive frequency $\omega$ for 500 periods of the modulation. We measure SR using $1/f^{\alpha}$ noise with $\alpha$ ranging from 0 to 2 in steps of 0.5. Using exponentially correlated noise, we measure SR with noise correlation times spanning nearly three orders of magnitude, from $\tau$ $\approx$ $0.01\Omega^{-1}$ to $\tau$ $\approx$ $10\Omega^{-1}$.

\section{Results}\label{results}

\subsection{1/f Noise}\label{1/f}

Trends in the SNR using $1/f$ noise are reported elsewhere \cite{Guerra-SR}. Specifically, results show that increasing the noise exponent suppresses the resonance in SNR, shifting it to higher noise powers and reducing the peak values reached. We note that interpretations of the underlying cause of these trends may vary, with an increasing effective potential \cite{Fuentes} and an increasing system asymmetry \cite{Wio} both serving as potential explanations.  Here we focus on alternative analyses that may help to illuminate the effects governing SR with $1/f$ noise. In these analyses, we draw heavily upon theoretical works which treat the canonical Duffing potential. Although the nonlinear behavior of the resonator is properly described by Eq.~\ref{duffing}, we note that application of the double-well model here is not strictly correct. The system under study is dynamic, with amplitude as the proper state variable (not position). Although the most accurate treatment would involve coexisting periodic attractors \cite{Dykman}, we focus on the canonical model in the interest of gaining some physical intuition for the effect of noise color on the system.

To begin, we apply a two-state filter to the recorded data and take the Fourier transform to examine the spectral response. Figure 2 displays the result of a typical measurement at low noise intensity. Several features present in this example arise consistently throughout the measurements taken with $1/f$ noise. Most prominently, the first few odd harmonics of $\Omega$ are present, decaying approximately as $1/n^{2}$, as expected. Also evident are even harmonics of the modulation frequency, which show little to no dependency on $n$. Finally, we note the presence of broad holes in the vicinity of even harmonics. The latter two observations are consistent with expectations for the weak noise limit \cite{Shneidman}. We note also that the presence of holes and/or peaks at even multiples of $\Omega$ have been a prominent feature of SR studies featuring asymmetric potentials \cite{Bartussek,Bulsara,Nikitin}. At higher noise powers, the holes $``$wash out$"$, disappearing as the overall noise level in the spectral response rises.

%------------------------------------------------------
%--------------Figure 2--------------------------------
%------------------------------------------------------
\begin{figure}
% Use the relevant command for your figure-insertion program
% to insert the figure file.
% For example, with the option graphics use
\resizebox{\columnwidth}{!}{%
  \includegraphics{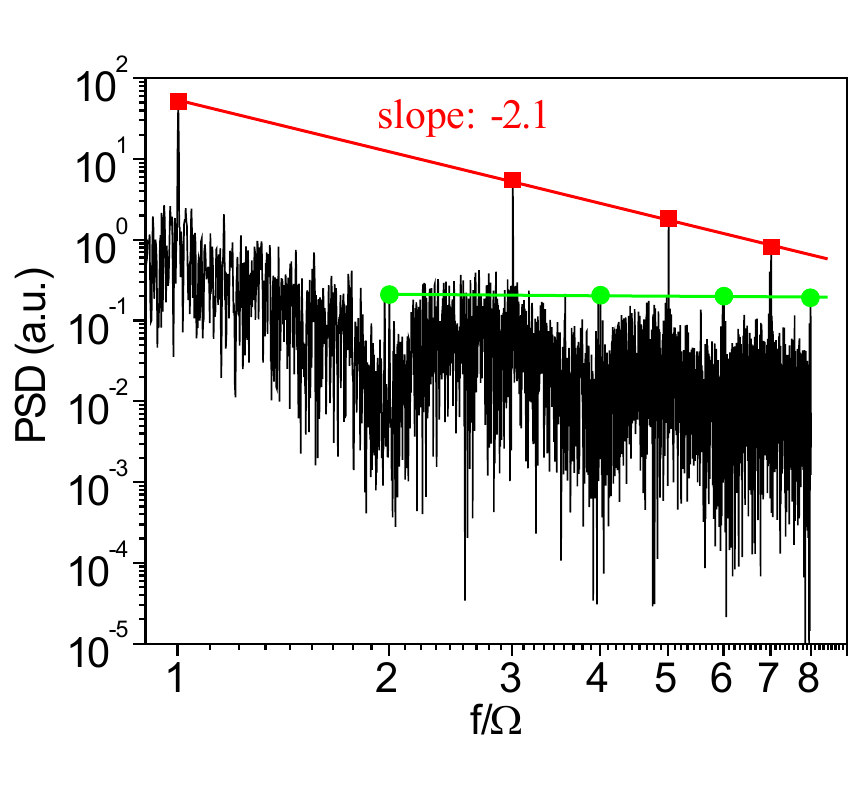} } 
\caption{An example of power spectral density for a single measurement. Here, $1/f^{\alpha}$ noise is used with $\alpha$ = 1 and $P_{N}$/$P_{D}$ = 0.01. Squares denote the odd harmonics and circles the even harmonics of the modulation frequency. Provided linear fits show that odd harmonics decay approximately as $1/n^{2}$ while even harmonics show no notable dependency with n.}
\label{psd}   
\end{figure}
%------------------------------------------------------
%------------------------------------------------------
%------------------------------------------------------

%------------------------------------------------------
%--------------Figure 3--------------------------------
%------------------------------------------------------
\begin{figure}
% Use the relevant command for your figure-insertion program
% to insert the figure file.
% For example, with the option graphics use
\resizebox{\columnwidth}{!}{%
  \includegraphics{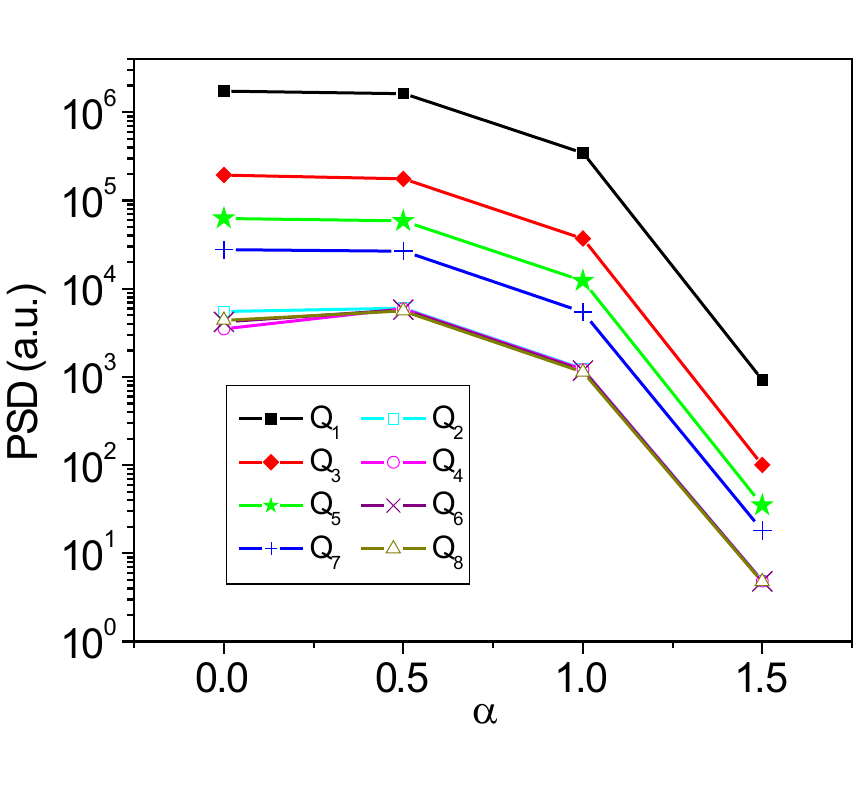} }
\caption{Dependency of the harmonics $Q_{n}$ of modulation frequency on $\alpha$ for a single noise power, $P_N$/$P_{D}$ = 0.0065. Here, we omit $\alpha$ = 2, as this noise spectrum proves insufficient to recover switching at this noise power.}
\label{qn}
\end{figure}
%------------------------------------------------------
%------------------------------------------------------
%------------------------------------------------------

%------------------------------------------------------
%--------------Figure 4--------------------------------
%------------------------------------------------------
\begin{figure}
% Use the relevant command for your figure-insertion program
% to insert the figure file.
% For example, with the option graphics use
\resizebox{\columnwidth}{!}{%
  \includegraphics{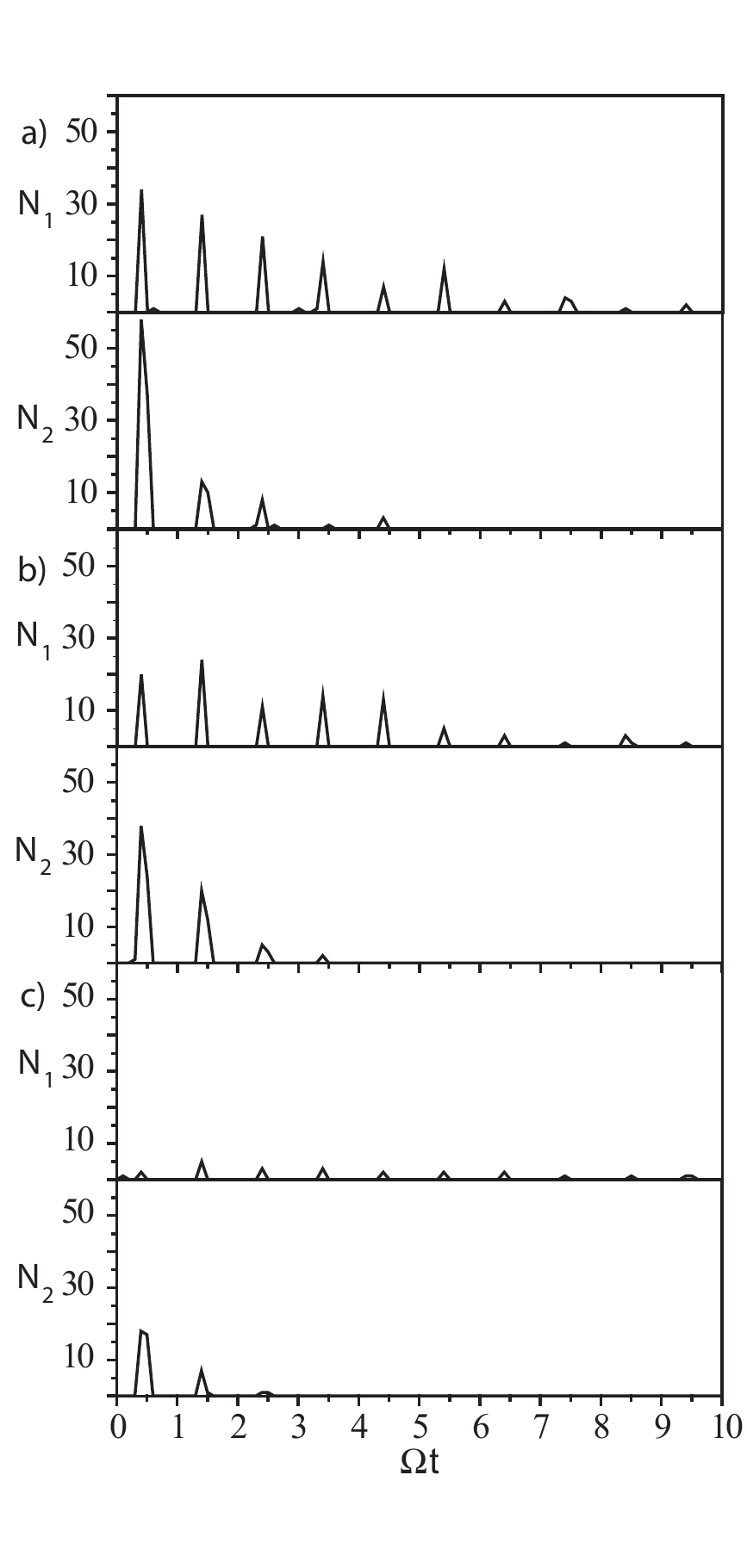} }
\caption{Example residence time distributions for the low and high amplitude states, denoted 1 and 2, respectively, at noise power $P_N$/$P_{D}$ = 0.005. Plots display data for: (a) $\alpha$ = 0, (b) $\alpha$ = 0.5 and (c) $\alpha$ = 1.}
\label{rtd}
\end{figure}
%------------------------------------------------------
%------------------------------------------------------
%------------------------------------------------------

In general, limitations on sampling rate make quantitative analysis of the harmonics $Q_{n}$ and the holes difficult and error prone; however, we note that at low noise powers, extrapolation of the linear fits shown in Fig.~\ref{psd} provides some evidence that the magnitudes of odd and even harmonics will cross sooner with higher $\alpha$.  Theoretical treatments predict oscillation in the relative magnitudes of odd and even $Q_{n}$ \cite{Nikitin}. As asymmetry increases, the magnitudes of the even harmonics surpass those of the odd harmonics at lower frequencies, the effect we see here with increasing $\alpha$. In addition, at low noise powers, odd harmonics decay monotonically with the noise exponent (Fig.~\ref{qn}). By contrast, even harmonics show a very slight maximum as a function of $\alpha$, an effect seen with increasing system asymmetry in both numerical simulations \cite{Bulsara} and theoretical calculations \cite{Nikitin}.

Residence time distributions help to provide a more complete picture of the system dynamics (Fig.~\ref{rtd}). At low noise powers, distributions consistently exhibit preferential occupation of one of the states, specifically the low amplitude one. Upon reaching the high amplitude state, the resonator frequently makes a correlated switch, returning to the low amplitude state after half a period of the modulation. By contrast, occupation of the low amplitude state may persist for multiple periods. In the plot shown for $\alpha$ = 1, for example, the resonator often remains in the low state for durations longer than those pictured, with the longest being for more than thirty periods of the modulation. Some asymmetry is also evident in measurements made with white noise, despite previously described efforts to symmetrize the potential (Sec.~\ref{meas}).

\begin{figure}
% Use the relevant command for your figure-insertion program
% to insert the figure file.
% For example, with the option graphics use
\resizebox{\columnwidth}{!}{%
  \includegraphics{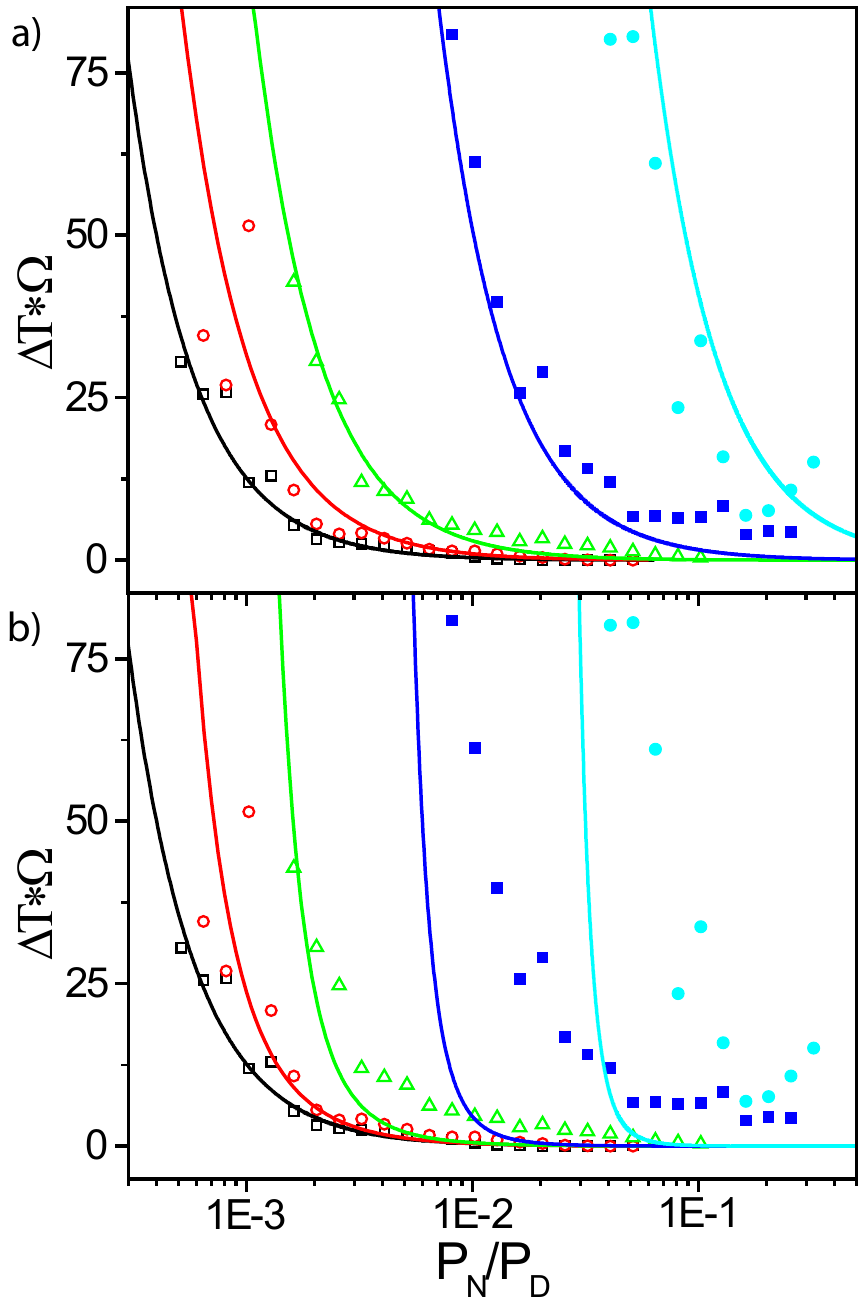} }
\caption{Difference $\Delta T$ between the average residence time of the two amplitude states vs. input noise power. The difference is expressed in units of the modulation period, 1/$\Omega$. The shapes denote the type of $1/f^{\alpha}$ noise used: $\alpha$ = 0 (empty squares), $\alpha$ = 0.5 (empty circles), $\alpha$ = 1 (empty triangles), $\alpha$ = 1.5 (filled squares), and $\alpha$ = 2 (filled circles). In both (a) and (b), the line for white noise represents the same two parameter fit to Eq.~\ref{eqdT}. Otherwise, solid lines in (a) and (b) depict fits with $c$ and $\Delta U$ allowed to vary, respectively, with the other variable held constant.}
\label{dT}
\end{figure}

To quantify any evolving asymmetry with noise color, we plot the difference $\Delta T$ between the average residence time of the two states vs. input noise power (Fig.~\ref{dT}). The effects of the asymmetry are most pronounced at low noise intensities, resulting in significant deviations between the observed average occupation times. In all cases, as noise power is increased and uncorrelated switching begins to dominate, the effects of asymmetry are mitigated, and the difference $\Delta T$ decays toward zero. Of note, however, is the shift in the $\Delta T$ vs. noise power data as we increase the noise exponent $\alpha$. Theoretical studies of switching time distributions predict the following form for $\Delta T$ with an asymmetry added to the Duffing potential \cite{Nikitin}:

\begin{equation}
\Delta T \propto \frac{ce^{\frac{\Delta U}{P_{N}}}}{P_{N}^{3/2}}
\label{eqdT}
\end{equation}

\noindent where $P_{N}$ is the noise power, $\Delta U$ is the potential barrier separating the states, and $c$ is the linear coefficient in the system potential, essentially determining the degree of asymmetry present. Here we express the noise power in units of the phase-modulated drive power $P_{D}$. Since increasing either or both of $c$ and $\Delta U$ could create the trend, discerning the cause of the observed shift is difficult. In two-parameter least squares fits (not pictured), the value of $c$ increases steadily with $\alpha$ while $\Delta U$ exhibits no clear trend; however, in some cases, initial parameter choices affect the final values. 

For comparison, we also attempt two separate sets of one-parameter fits to Equation~\ref{eqdT}. With $\Delta U$ fixed at its value from the white noise ($\alpha$ = 0) two-parameter fit, Figure~\ref{dT}(a) displays regressions with $c$ as the fitting parameter; likewise, Figure~\ref{dT}(b) shows fits with $\Delta U$ allowed to vary and $c$ fixed at its value from the white noise data. Clearly the former set better captures the trend seen with increasing $\alpha$. In essence, a steady increase in asymmetry is necessary to account for the severe shift observed. We conclude that increasing noise color intensifies the system asymmetry, though we cannot rule out the possibility that the effective potential barrier deepens as well.

%------------------------------------------------------
%--------------Figure 5--------------------------------
%------------------------------------------------------
%------------------------------------------------------
%------------------------------------------------------
%------------------------------------------------------

\subsection{Ornstein-Uhlenbeck Noise}\label{OU}

\begin{figure}
% Use the relevant command for your figure-insertion program
% to insert the figure file.
% For example, with the option graphics use
\resizebox{\columnwidth}{!}{%
  \includegraphics{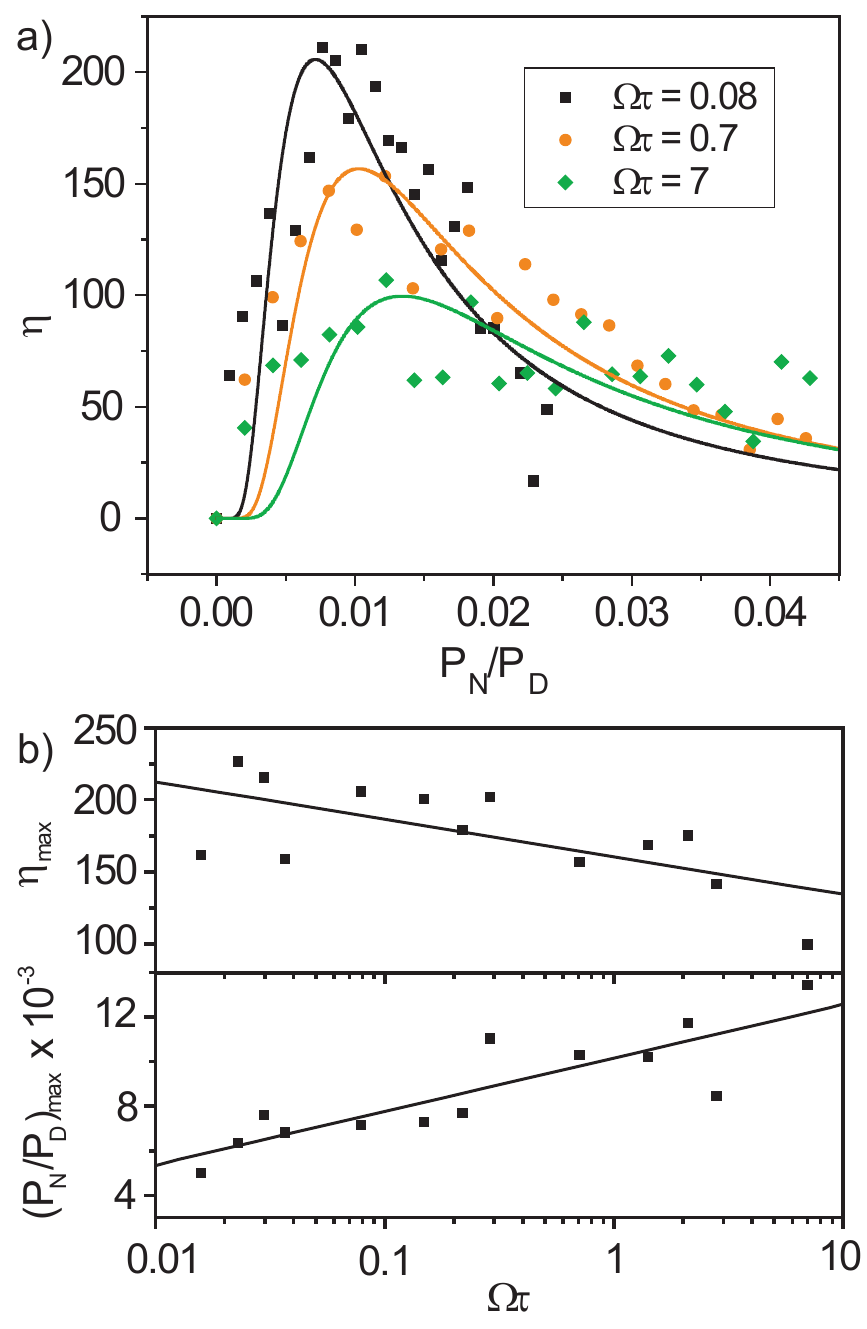} } 
\caption{a) Spectral amplification vs. noise power using Ornstein-Uhlenbeck noise with three exemplary values of correlation time $\tau$. Additional values of $\tau$ are omitted for clarity. Solid lines depict fits to Eq.~\ref{ucna}. Correlation times are measured using a spectrum analyzer and are presented here as a unitless number scaled by the modulation frequency (i.e. $1/\Omega$ = 1).  b) Maximum spectral amplification and corresponding noise power necessary to reach that maximum, respectively, vs. noise correlation time. Values are taken from the fits described in (a). Solid lines are included as a guide to the eye and are not intended as fits to any theoretical form.}
\label{lorentzian}
\end{figure}

Extending our measurements to another class of colored noise, we also examine the effect of correlation time on stochastic resonance with the use of exponentially correlated noise. Noise power $P_N$ is defined as the power contained within the FWHM of the resonance, and it is again expressed in units of the phase-modulated drive power $P_D$. Spectral amplification, defined as $S(\Omega)|_{P_N}/S(\Omega)|_{P_{N}=0}$, \\ where $S(\Omega)$ is the PSD at the modulation frequency, is monotonically suppressed as the correlation time of the noise spectrum becomes longer (Fig~\ref{lorentzian}). Peak amplification decreases and shifts to higher noise powers as the noise color increases.

Since it is more easily approached mathematically than $1/f$ noise, Ornstein-Uhlenbeck noise is more prevalent in theoretical treatments of SR. In particular, it has been explored via the Unified Colored Noise Approximation (UCNA) \cite{Hanggi}. Our measurements in the nonlinear regime are beyond the applicability of such treatments, but as a means of emphasizing the trends, we fit with two parameters to the basic form predicted for spectral amplification $\eta$:

\begin{equation}
\eta = (\frac{A}{P_N/P_D})^2\frac{1}{1+e^{(\frac{B}{P_N/P_D})}}
\label{ucna}
\end{equation}

\noindent Solid lines in Figure~\ref{lorentzian}(a) depict these fits, from which the peak value of $\eta$ and the corresponding noise power at the peak are extracted and plotted (Fig.~\ref{lorentzian}(b)). Over almost three orders of magnitude in $\tau$ we observe a very slight shift in peak location, a notable contrast with the nearly linear shift predicted by linear response theory \cite{Hanggi} and with the much more significant shift seen when increasing $1/f$ noise color \cite{Guerra-SR}.

%------------------------------------------------------
%--------------Figure 6--------------------------------
%------------------------------------------------------
%------------------------------------------------------
%------------------------------------------------------
%------------------------------------------------------

\section{Discussion}\label{disc}

Stochastic resonance using $1/f$ noise has been studied experimentally in electronic \cite{Kiss} and biological systems \cite{Nozaki}, and it has been the focus of some theoretical work \cite{Fuentes}. Of particular note are parallels to measurements made with a Schmitt trigger, in which even harmonics and holes were evident in the PSD, and the $1/f$ noise was interpreted as creating a slowly-varying asymmetry in the system potential \cite{Kiss}. Here we extend the approach by varying the value of the noise exponent $\alpha$ and find evidence that increasing the noise color exacerbates the asymmetry, an intuitive result. As a consequence of the increasing asymmetry, the resonance in signal-to-noise ratio should shrink and shift to higher noise powers \cite{Wio}, as observed elsewhere \cite{Guerra-SR}.

Using exponentially correlated noise, the suppression of stochastic resonance we observe with increasing correlation time agrees well with theoretical predictions \cite{Hanggi}, analog simulations \cite{Gaimmatoni-PRL/PRA} and experimental results obtained in a tunnel diode \cite{Mantega}. Measurement in the nonlinear regime separates these results slightly from those of other investigations, however, the conclusion that the use of colored noise reduces the effectiveness of SR is a common one. To optimize the performance of the nanomechanical resonator as a two-state system, white noise proves superior to both $1/f$ and Ornstein-Uhlenbeck noise types, inducing greater peak amplification at smaller noise powers.

Analyzing the results with Ornstein-Uhlenbeck noise for trends in asymmetry, we find no evidence of a connection similar to that seen with $1/f$ noise, a difference that may stem from our placement of the noise spectra in frequency space (Sec.~\ref{meas}). Even harmonics and holes in the PSD again persist for small noise intensities and disappear as the weak noise limit is eclipsed; however, plots of $\Delta T$ vs. noise power exhibit no measurable shift with changing correlation time. This result, combined with the lack of a significant shift of the spectral amplification resonance to higher noise powers, suggests no strong effect of the noise correlation time on potential depth or asymmetry for this system.

\section{Conclusion}\label{conc}

Stochastic resonance is measured in a bistable silicon nanomechanical resonator using a wide range of colored noise spectra. Evidence suggests that escalation of $1/f^{\alpha}$ noise color increases an inherent potential asymmetry in the nanomechanical two-state system, explaining the suppression of signal-to-noise ratio seen with this noise class. Increasing correlation time with Ornstein-Uhlenbeck noise monotonically suppresses the stochastic resonance, though the shift observed in resonance over three orders of magnitude in $\tau$ is substantially less than that observed for $1/f$ noise when changing $\alpha$ from 0 to 2. In either case, increasing the color of the noise spectrum weakens the effectiveness of stochastic resonance, making white noise the best option for optimizing device performance.

\acknowledgement{This work is supported by NSF (DMR-0449670).}

\end{document}